\documentclass[referee]{aastex631}

\usepackage{graphics}
\begin{document}

\title{A unified model of solar prominence formation with self-consistent heating}

\author{C. J. Huang}
\affiliation{School of Astronomy and Space Science, Nanjing University, Nanjing 210023, PR China}
\affiliation{Key Laboratory for Modern Astronomy and Astrophysics, Nanjing University, Nanjing 210023, PR China}
\affiliation{Department of Astronomy, Kyoto University, Kyoto 606-8502, Japan}

\author[0000-0002-9908-291X]{Y. W. Ni}
\affiliation{School of Astronomy and Space Science, Nanjing University, Nanjing 210023, PR China}
\affiliation{Key Laboratory for Modern Astronomy and Astrophysics, Nanjing University, Nanjing 210023, PR China}

\author[0000-0002-4205-5566]{J. H. Guo}
\affiliation{School of Astronomy and Space Science, Nanjing University, Nanjing 210023, PR China}
\affiliation{Key Laboratory for Modern Astronomy and Astrophysics, Nanjing University, Nanjing 210023, PR China}

\author[0000-0002-7289-642X]{P. F. Chen}
\affiliation{School of Astronomy and Space Science, Nanjing University, Nanjing 210023, PR China}
\affiliation{Key Laboratory for Modern Astronomy and Astrophysics, Nanjing University, Nanjing 210023, PR China}
\affiliation{State Key Laboratory of Lunar and Planetary Sciences, Macau University of Science and Technology, Macau 999078, PR China}

\correspondingauthor{P. F. Chen}
\email{chenpf@nju.edu.cn}

\begin{abstract}
Several models have been proposed to explain the formation of solar prominences, among which the evaporation--condensation model and the direct injection model are the most popular ones. In our previous study we proposed to unify these two models, namely, both are due to localized heating in the chromosphere, presumably via magnetic reconnection. When the localized heating is located in the upper chromosphere, the cold in-situ plasmas are heated to coronal temperatures, then evaporated to the corona, and finally condensate to form a prominence. Such a process is manifested as the evaporation-condensation model. When the localized heating is located in the lower chromosphere, the enhanced in-situ pressure  would push the cold plasmas in the upper chromosphere to the corona directly, which is manifested as the direct injection model. While the idea was confirmed by the one-dimensional hydrodynamic simulations, the heating was imposed ad hoc. In order to simulate the localized heating more self-consistently, we perform two-dimensional magnetohydrodynamic simulations in this paper, where the localized heating is naturally realized by magnetic reconnection at different heights. The simulations further validate our model. Besides, mass circulation in the solar atmosphere is also briefly discussed.
\end{abstract}

\keywords{Solar prominences (1519) --- Solar filaments (1495) --- Magnetohydrodynamic simulations (1966)}

\section{Introduction} \label{sec:intro}

Solar prominences are cold ($\sim 7\times 10^3$ K) and dense ($\sim 10^{10}$--$10^{11}$ cm$^{-3}$) plasmas suspended in the hot and tenuous corona above photospheric magnetic polarity inversion lines \citep{Labr10}. They appear bright above the solar limb and dark against the solar disk, in the latter of which they are called solar filaments. They have attracted wide attention for their intriguing characteristics in each stage of their lifetime from formation to eruption \citep{Chen20}. The formation of prominences is related to thermal instability or thermal nonequilibrium \citep{Anti91, Klim19a, Klim19b}, their existence is accompanied by ceaseless counterstreamings and occasional global oscillations \citep{Zirk98, Jing03, Zhan13, Zhan19, Zhou18, Lido18}, as well as mass circulation \citep{Xiac16}, and their eruptions are intimately related to solar flares and coronal mass ejections \citep[CMEs,][]{Jing04, Chen11, purk24}.

The formation mechanisms of solar prominences have been investigated for decades either through observations \citep{Chae00, Chae03, Berg12, Liuw12, Wang18, Lile19, Wang19, Vial20, Xias23} or via numerical simulations \citep{Wust90, Anti91, Anti99, Dahl98, Karp05, Karp08, Xiac11, Xiac12, Xiac16, Jerc24, Donn24, Yosh25}. Several models have been proposed, among which the evaporation--condensation model and the direct injection model are the two most popular ones \citep{Mack10}. In the evaporation–condensation model, heated chromospheric plasma evaporates into the corona, where it undergoes catastrophic cooling caused by thermal instability or thermal nonequilibrium, and condensates to form a prominence. This model can naturally explain the sudden appearance of cold filaments observed in the H$\alpha$ images. In the direct injection model \citep[e.g.,][]{Wang19}, cold plasmas in the chromosphere are ejected into the corona along magnetic field lines dynamically. The injection process is observed as jet flows of cold plasma. The acceleration of the cold jets is presumably provided by Lorentz force, but their dynamics is significantly influenced by gravity and gas pressure. The cold materials would fall back to their original site, or overshoot to the other footpoint of the magnetic field line, forming a dynamic filament with short-lived threads, or be trapped in the magnetic dip by gravity, forming a rather stable solar filament.

With one-dimensional (1D) hydrodynamic simulations, we found that the two most popular models can be unified \citep{Huan21}, i.e., both models are related to localized heating in the chromosphere and the main difference leading to the bifurcation is the height of the heating. If the localized heating lies in the upper chromosphere, the local cold plasma is heated to several million Kelvin, leading to evaporation into the corona. As the coronal density near magnetic dips increases, in-situ coronal condensation happens, presumably due to thermal instability rather than thermal nonequilibrium as the latter relates to the fact that no true stationary state can be established with the exact balance between heating and cooling. Such a process is in accord with the evaporation--condensation model. In contrast, if the localized heating is situated in the lower chromosphere, the cold plasma is heated, which enhances the gas pressure and pushes the cold plasma in the upper chromosphere to move upward, forming an upflow injected into the corona. Such a process resembles the direct injection model. Such a unified model was confirmed unintentionally by \citet{kuce24}.

It is noted however that in the 1D model, the localized heating was realized by simply depositing energy in a local region, without considering what is in charge of the localized heating. In observations, there is mounting evidence indicating that chromospheric magnetic reconnection is the driver for filament formation \citep{Chae00, Wang01, Chae03, Zoup16, Yanx16, Yang19}. Hence, it would be of great interest to extend our 1D hydrodynamic simulations to two-dimensional (2D) magnetohydrodynamic (MHD) simulations so that we can check whether magnetic reconnection at different heights of the solar chromosphere can indeed reproduce the filament formation described by the two traditional models. This paper is organized as follows. The numerical method is introduced in \S\ref{sec2}, and the simulation results are presented in \S\ref{sec3}, which are followed by discussions in \S\ref{sec4}.

\section{Numerical method} \label{sec2}

\subsection{Governing equations}

The full set of 2D MHD equations as shown below, with empirical heating, thermal conduction, radiative cooling, and anomalous resistivity included, are solved by the open-source code MPI-AMRVAC \citep{Xia2018, Keppens2023}.

\begin{equation}
\begin{array}{cc}
\frac{\partial \rho}{\partial t}+\nabla\cdot(\rho {\bf v})=0, \\
\frac{\partial {\rho \bf v}}{\partial t}+\nabla \cdot (\rho \bf{vv}+\it{p}_{\rm tot}\bf{I}-\frac{\bf{BB}}{\mu_{0}})=\rho \bf{g}, \\
\frac{\partial e}{\partial t}+\nabla \cdot (e\bf{v}-\frac{\bf{BB}}{\mu_{0}}\cdot \bf{v}+\bf{v}\it{p}_{\rm{tot}})=\rho\bf{g}\cdot\bf{v}+\nabla\cdot(\bf{\kappa} \cdot \nabla \it{T})-n_{\rm H}n_{\rm e} \rm{\Lambda}(\it{T})+\it{H}+\nabla\cdot(\bf{B}\times \eta\bf{J}), \\
\frac{\partial \bf{B}}{\partial t}+\nabla\cdot(\bf{vB}-\bf{Bv})=-\nabla\times(\eta\bf{J})
 \\
\end{array}
\end{equation}
where $\rho = 1.4m_{\rm p}n_{\rm H}$ is the mass density with the consideration of the helium abundance $\omega_{\rm He}=0.1$. $n_{\rm H}$ is the number density of the proton and $m_{\rm p}=1.67\times 10^{-24}\ \rm g$. $\mathbf v$, $\mathbf B$, and $\mathbf J$ are the velocity, magnetic strength, and current density. $p = 2.3n_{\rm H}k_{\rm B}T$ is the gas pressure, where $k_{\rm B}$ is the Boltzmann constant. $e = \rho v^2 /2 + p/(\gamma -1) +B^2/2\mu_{0}$ is the total energy density, and $\gamma =5/3$ is the adiabatic index. $\mathbf{g}=-g_{\sun}\bf{e_y}$ is the gravitational acceleration, and $g_{\sun} = 274$ m s$^{-2}$. $\kappa=10^{-6}T^{5/2}\ \rm erg\ cm^{-1}\ s^{-1}\ K^{-1}$ is the Spitzer heating conductivity. $n_{\rm H}n_{\rm e}\rm{\Lambda}(\it{T})$ is the optically-thin cooling term. There are several models to choose for the cooling term. Here we choose the JCcorona model, which is based on \citet{Colg08} but modified by \citep{Xiac11}. According to \citet{herm21}, such a model leads to a slightly faster condensation. $H$ is the steady background heating, aiming at balancing the radiative cooling so as to maintain a hot corona. It is set as below:

\begin{equation}
H(y)=H_{0}e^{-y/h_{0}},
\end{equation}
where $H_0=7.0\times 10^{-5}\ \rm erg\ cm^{-3}\ s^{-1}$ and the scale height $h_{0}=35\ \rm Mm$. Different forms of the heating term would result in slightly different atmosphere. Our parameters are chosen so that the resulting chromosphere has a thickness of $\sim$1.5 Mm as indicated by observations. The background heating is independent of time and magnetic field. We assume the anomalous resistivity as below to induce fast reconnection at the left footpoint of the magnetic loops:

\begin{equation}
\eta (\xi)=\left\{
\begin{array}{cc}
     \eta_{0}  \min(1,|\frac{\xi}{\xi_{m}}|),& |\xi|\ge \xi_c,  \\
     0, &  |\xi|<\xi_c,
\end{array}
         \right.
\end{equation}
where $\xi = J_z/|\mathbf{B}|$, $\xi_c = 4$ represents the critical threshold above which anomalous resistivity is activated, $\xi_m$ is the threshold above which the resistivity saturates, and $\eta_{0}$ denotes the maximum resistivity.

\subsection{Initial and boundary conditions}

\begin{figure}
    \centering
  \includegraphics[width=18cm]{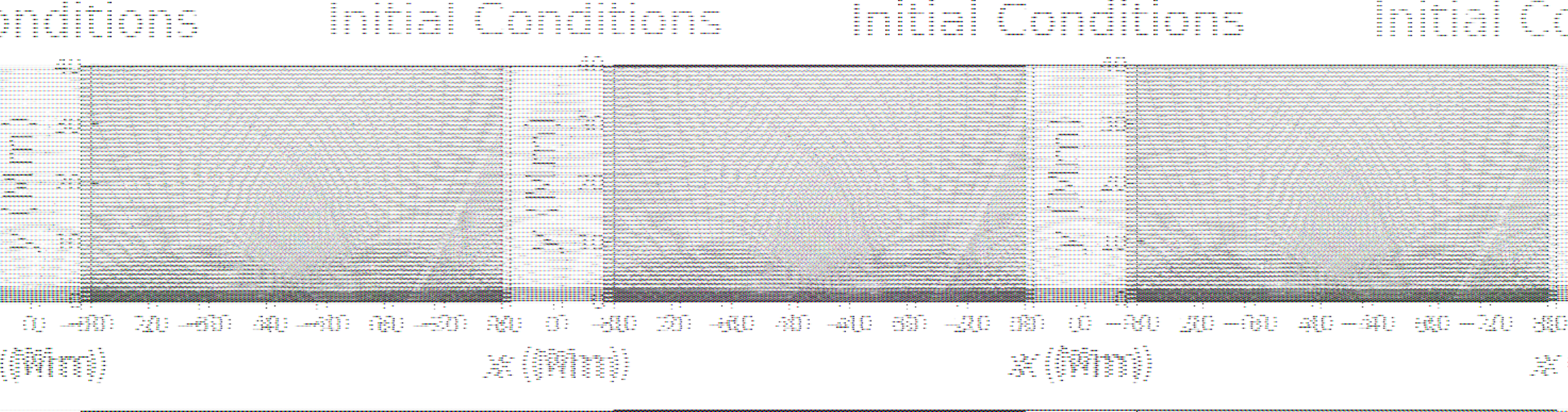}

    \caption{Left panel: Initial magnetic field ({\it solid lines}) overlaid on the density distribution ({\it color scale}); Right panel: Initial magnetic field ({\it solid lines}) overlaid on the temperature distribution ({\it color scale}).}
    \label{fig1}
\end{figure}

The initial thermal distribution is taken to mimic the quiet solar atmosphere, with the consideration of cold chromosphere and hot corona. Following previous works \citep{Terr16, Zhou18, Zhan19}, we set the temperature distribution as follows.

\begin{equation}
T(y)=\left\{
\begin{array}{cc}
	T_{\rm cho}+(T_{\rm co}-T_{\rm cho}) (1+\tanh((y-y_{\rm tr}-c)/w_{\rm tr})/2), & y \le y_{\rm tr},  \\
     \left[3.5F_{\rm c}(y-y_{\rm tr})/\kappa_{c}+T_{\rm tr}^{3.5}\right]^{2/7}, & y > y_{\rm tr},
\end{array}
      \right.
\end{equation}

\noindent
where $T_{\rm cho}=8000$ K, $T_{\rm tr}=1.6\times 10^5$ K, and $T_{\rm co}=1.5\times 10^6\ $K are the temperatures at the bottom of the low chromosphere, top of the transition region, and top of the corona in our numerical domain. $y_{\rm tr}$=2 Mm is the initial height of the transition region. $F_{c}=2\times 10^{5}\ \rm erg\ cm^{-2}\ s^{-1}$ is the constant vertical thermal conduction flux and $\kappa_{c} = 8 \times 10^{-7}~\mathrm{erg\,cm^{-1}\,s^{-1}\,K^{-1}}$ is the Spitzer type heat conductivity. We set the width of the transition region $w_{\rm tr}$= 0.2 Mm and $c=0.27$ Mm to extend the temperature profile from the corona to the chromosphere.

We use four line currents below the bottom boundary to generate a potential magnetic field with magnetic dips in the simulation domain, which is expressed as

\begin{equation}
\begin{array}{lc}
     B_{x}(x,y)/B_{0}=+[(y-y_{1})/r_{1}^{2} + (y-y_{4})/r_{4}^{2}]L_0-[(y-y_{2})/r_{2}^{2} + (y-y_{3})/r_{3}^{2}]L_0,  \\
     B_{y}(x,y)/B_{0}=-[(x-x_{1})/r_{1}^{2} + (x-x_{4})/r_{4}^{2}]L_0+[(x-x_{2})/r_{2}^{2} + (x-x_{3})/r_{3}^{2}]L_0,  \\
\end{array}
\end{equation}
where $B_0=400$ G, $x_1=-x_4=-60\ \mathrm{Mm}$, $x_2=-x_3=-45\ \mathrm{Mm}$,
$y_1=y_4=-25\ \mathrm{Mm},$ $y_2=y_3=-17.5\ \mathrm{Mm}$, $r_i=\sqrt{(x-x_i)^2+(y-y_i)^2},~i$=1, 2, 3, and 4, and $L_0=10$ Mm is the characteristic length in nondimensionalize the space in the MHD equations. As seen in Figure \ref{fig1}, the magnetic field is symmetric with the maximum strength being 65.3 G at the location $(\pm 5.2\ \mathrm{Mm},~0)$. As a reference, the plasma $\beta$, the ratio of gas to magnetic pressure, is 0.013 at the position (0, 10) Mm, which is typical in the corona.

The simulation box has the size of 160 Mm $\times$ 40 Mm. The base-level grid mesh consists of 80 $\times$ 80 grid points, and 5-level refinement is applied. The effective resolution is 125 km $\times$ 31.25 km. We fix all the physical variables on the bottom, left, and right boundaries, whereas the top boundary is an open one. With the initial atmosphere and magnetic field imposed, we perform MHD simulations to relax the initial conditions to a quasi-static state. In the previous 1D hydrodynamic simulations of our unified model \citep{Huan21}, the height and width of the transition region barely change during the relaxation period. However, in the 2-dimensional MHD cases, both thermal and magnetic distributions evolve to new equilibria. Figure \ref{fig1} depicts the magnetic field overlaid on the density distribution (left panel) and the temperature distribution (right panel). It is noticed that the region near the magnetic separatrix is heated to a higher temperature, and the transition region changes as well, though slightly. Such a new equilibrium is used as the real initial condition for our research.

\begin{figure}
    \centering
    \includegraphics[width=0.8\textwidth]{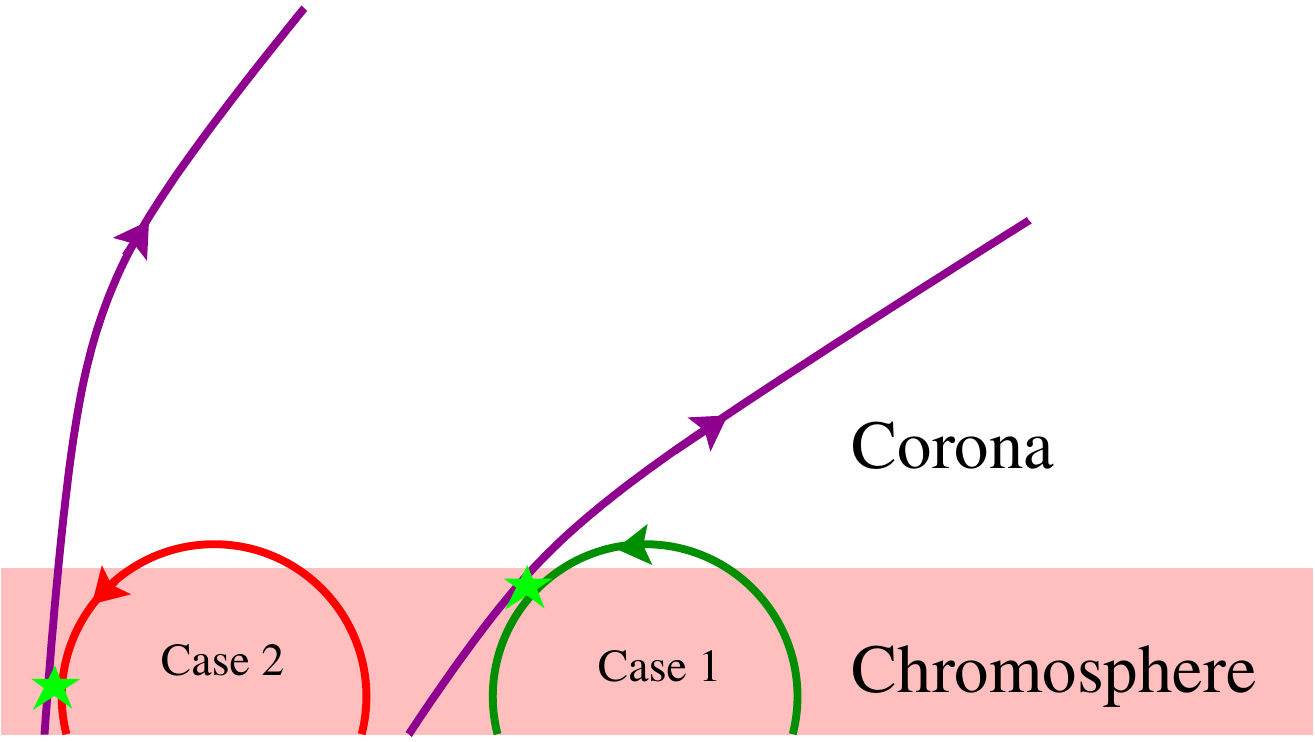}
    \caption{Schematic sketch of the interaction between ephemeral flux and the left footpoint of the preexisting magnetic field. The green line marks the location of the ephemeral magnetic field in Case 1, and the red line marks that of the ephemeral flux in Case 2. The asterisks indicate the contact points between the ephemeral flux and the anti-parallel preexisting magnetic field.}
    \label{fig2}
\end{figure}

In order to simulate the localized chromospheric heating in a natural way, we impose an ephemeral bipolar magnetic field near the left footpoints of dipped magnetic field lines, as done by \citet{Chen00}. The strength is chosen so that the shallow magnetic dips would not be destroyed and the prominences can be maintained. The ephemeral field is described by ${\bf B}_{\rm eph}(x,~y,~t)=A(t){\bf B}^s_{\rm eph}(x,~y)$ with the spatial distribution similar to that in the previous works \citep{Jian12} at the bottom boundary, i.e.,

\begin{equation}
A(t)=\left\{
\begin{array}{cc}
B_{\rm 0}t/t_{\rm em},     & t \leq t_{\rm em},\\
	B_{\rm 0}[1-(t-t_{\rm em})/t_{\rm sub}],   & t_{\rm em}<t<t_{\rm em}+t_{\rm sub}, \\
0, & t \ge t_{\rm em}+t_{\rm sub},
\end{array}
\right.
\end{equation}

\begin{equation}
	\bf{B}^s_{\rm eph}(x,y)=\left\{
\begin{array}{cc}
    [(y-y_{\rm e})/r^2 \vec{e_x}- (x-x_{\rm e})/r^2 \vec{e_y}]L_0, & r < r_{\rm e},\\
    0, & r > r_{\rm e},
\end{array}
\right.
\end{equation}
where $B_{\rm 0} = 13\ \rm G$  determines the strength of the ephemeral field, $t_{\rm em}$ is the flux emerging time, $t_{\rm sub}$  is the submerging time, $x_{\rm e}$ and $y_{\rm e}$ are the coordinates of the ephemeral line current, $r=\sqrt{(x-x_e)^2+(y-y_e)^2}$, and $r_e=1.25$ Mm. In this paper, $y_e$ is fixed to be $- 0.25$ Mm. Owing to the inhomogeneity of the magnetic field orientation, the highest contact point between the ephemeral flux and the antiparallel preexisting field is at different heights. As illustrated by Figure \ref{fig2}, the corresponding magnetic reconnection occurs at the upper chromosphere in Case 1 and at the lower chromosphere in Case 2.

\section{Results} \label{sec3}

\begin{figure}[htbp]
    \centering
    \includegraphics[width=0.8\textwidth]{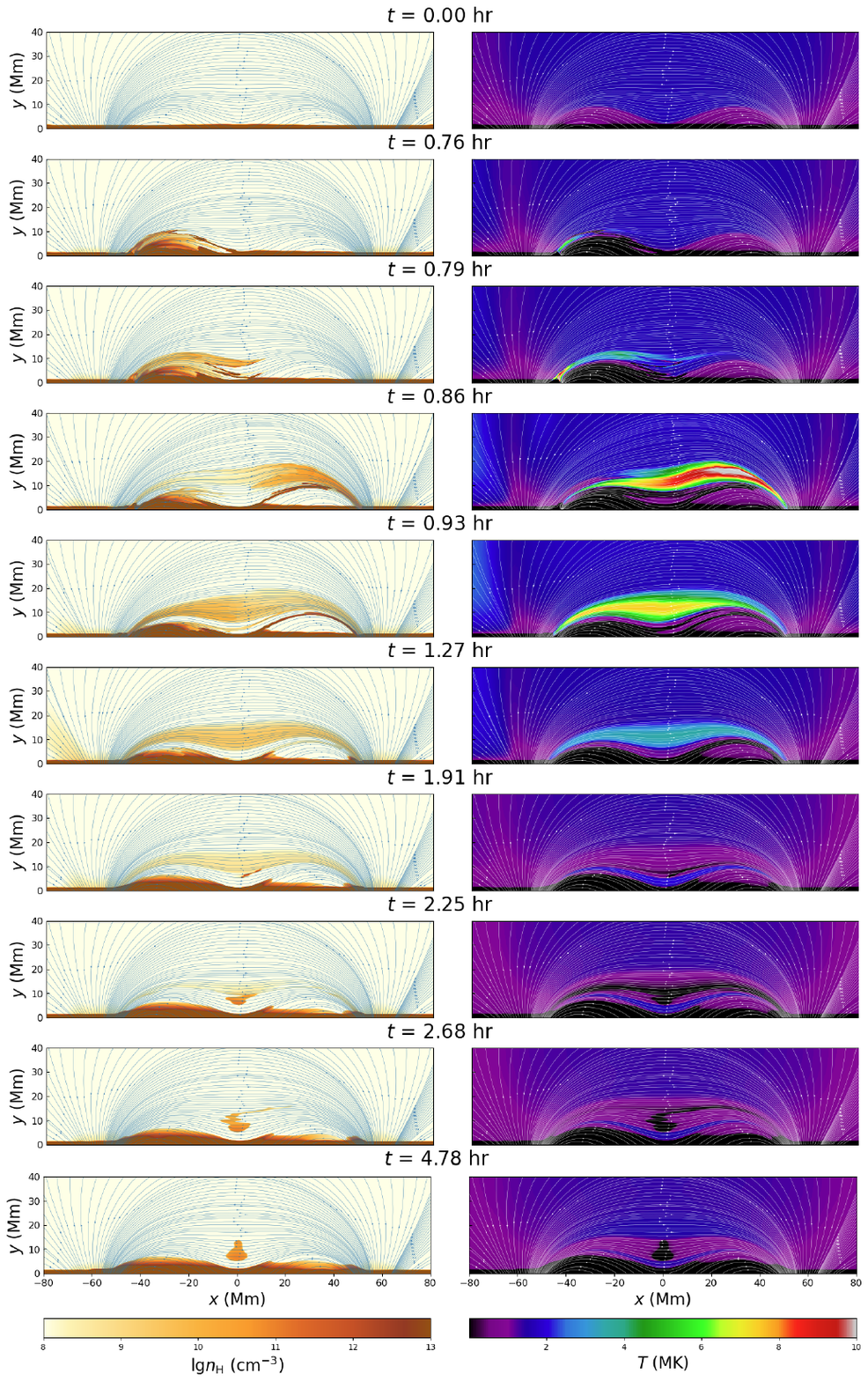}
    \caption{Evolution of magnetic field, density, and temperature in Case 1. The left column showcases the magnetic field ({\it solid lines}) and density ({\it color scale}), whereas the right column shows the magnetic field ({\it solid lines}) and temperature ({\it color scale}). This case resembles the evaporation--condensation model. Animation of this figure is available online.}
    \label{fig3}
\end{figure}

In Case 1, the ephemeral magnetic flux is centered at $x_{\rm e} = -44\ \rm{Mm}$, with $t_{\rm em} = 3010\ \rm{s}$ and $t_{\rm sub} = 430\ \rm{s}$. For the anomalous resistivity, we set $\xi_m = 15$ and $\eta_{0} = 0.25$. The evolution of the number density and temperature distributions is shown in the left and right columns of Figure \ref{fig3}, respectively. As the ephemeral flux emerges from the bottom boundary, it interacts with the preexisting magnetic field in the chromosphere, creating a magnetic null point. This interaction initially drives a slight upward displacement of the transition region.

Before the impulsive ejection driven by magnetic reconnection, an episode of cold plasma is ejected due to melon-seed effect, this weak cold jet moves beneath the left shoulder of the magnetic loop and falls back down to the chromosphere after a short period of time, rather than entering the magnetic dip. As the emergence proceeds, the magnetic null point continues to rise. Subsequently, magnetic reconnection is triggered and a cold plasma surge is ejected, as seen at $t$=0.76 hr. The cold surge travels along the magnetic field line and eventually impacts the solar surface at the right footpoint of the magnetic loop. The draining material collides with the chromosphere at the right footpoint, generating heating and reflected mass flow. Meanwhile, hot plasma outflow produced by the magnetic reconnection is supplied to the corona continuously, as seen at $t$=0.79 hr and $t$=0.86 hr. Due to the magnetic frozen-in effect, the trajectory of the fast hot flow is constrained to the magnetic field line. As a result, the compression at the right-hand side of the magnetic dip and the right shoulder leads to two additional episodes of heating, which are responsible for the transient appearance of the high-temperature plasma.

As the ephemeral magnetic field begins to submerge at $t$=0.83 hr, magnetic reconnection gradually fades away and extinguishes eventually. Thereafter, under the influence of the radiative cooling, which becomes very effective due to the high density in the magnetic dip, the temperature of the trapped plasma starts to fall down as seen at $t$=1.27 hr. Finally, at $t$=1.91 hr, the coronal condensation is triggered at the lower part of the magnetic dip and starts to grow above the PIL. After $t$=2.25 hr, additional condensations form near both shoulders and subsequently fall back to the chromosphere, manifesting as coronal rains. Due to slight variations in the onset locations and evolution of condensation across different magnetic field lines, the prominence exhibits small-amplitude decaying oscillations before reaching a quasi-stable state. The oscillation period is consistent with the pendulum model as demonstrated by \citet{Zhou18}. Unlike \citet{Jerc23} who imposed intermittent heating at the chromosphere so that their prominence oscillation is dampless, the prominence in our simulations becomes largely stabilized by $t$=4.78 hr. At this time, the prominence has a vertical extent of approximately 8.12 Mm above the solar surface and a thread length (i.e., the maximum extent along the direction parallel to the solar surface) of about 8.48 Mm. In this case, the prominence exhibits a minimum temperature of $\sim1.9 \times 10^4\ \rm K$ and a maximum density up to $10^{9.8}\ \rm cm^{-3}$.

\begin{figure}[htbp]
    \centering
    \includegraphics[width=0.9\textwidth]{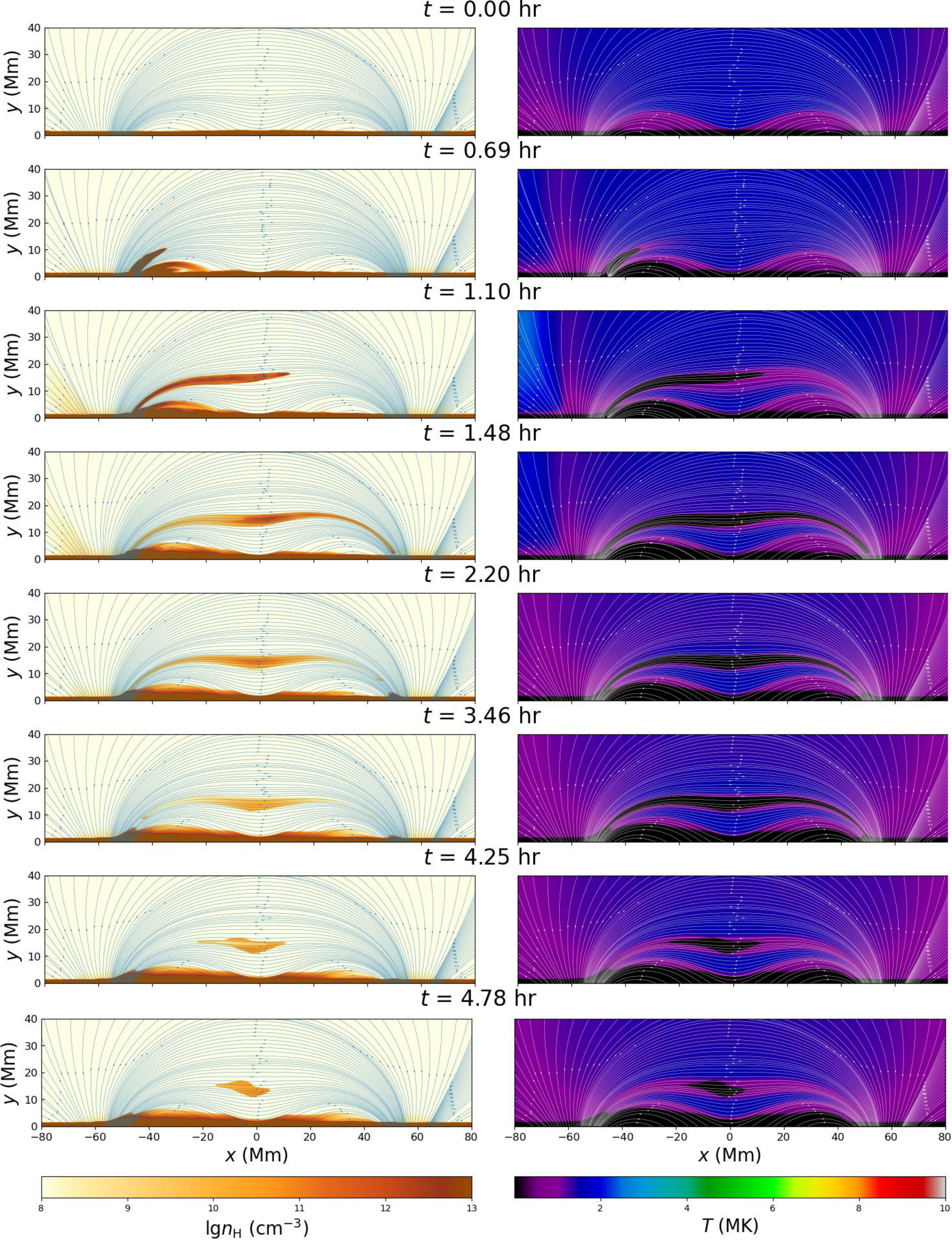}
    \caption{Evolution of magnetic field, temperature, and density in Case 2. The left column showcases the magnetic field ({\it solid lines}) and density ({\it color scale}), whereas the right column shows the magnetic field ({\it solid lines}) and temperature ({\it color scale}). This case resembles the direct injection model. Animation of this figure is available online.}
     \label{fig4}
\end{figure}

 With all other parameters being the same, the ephemeral magnetic field is shifted $3\ \rm Mm$ to the left in Case 2, i.e., the ephemeral flux is centered at $x_{\rm e}=-47$ Mm, with a shorter $t_{\rm em} = \rm 2580\ s$ and a longer $t_{\rm sub} = \rm 1720\ s$. The anomalous resistivity is adjusted to $\xi_m = 40$ and $\eta_{0} = 0.08$. Since the local preexisting magnetic field is more vertical near the new ephemeral site, the contact point between the ephemeral flux and the preexisting field becomes lower, as illustrated in Figure \ref{fig2}. Consequently, magnetic reconnection occurs in the lower solar chromosphere.

The evolution of plasma density and temperature in Case 2 is depicted in the left and right panels of Figure \ref{fig4}, respectively, where the magnetic field lines are overlaid on. As observed at $t$=0.69 hr, a small ejection of cold plasma driven by the melon-seed effect, which also occurs in Case 1, appears near the left footpoint. As the ephemeral flux continues to emerge, magnetic reconnection is triggered, accelerating a parcel of cold plasma upward initially via the magnetic Lorentz force, and subsequently by gas pressure as the reconnected field lines straighten. The injected cold flow is ejected with an average velocity of $\sim$40 $\rm km\ s^{-1}$.

Because the ephemeral flux is shifted leftward, the cold jet travels along magnetic field lines with higher and shallower dips. The shallow geometry of these dips results in only a weak component of gravitational force along the field lines. As a result, the upper part of the cold jet overshoots the right shoulder of the magnetic loop and eventually drains back to the chromosphere, as seen at $t$=1.48 hr. After the ephemeral field disappears and no further cold material is ejected, the remaining cold plasma at both shoulders either falls back to the chromosphere or settles into the magnetic dip, driven by gravity.  The prominence stabilizes 2.58 hours later at $t$=4.78 hr, with the maximum density and minimum temperature similar to those in Case 1. The vertical extent of the prominence is 6.4 Mm and the longest thread is $\sim$16.7 Mm.

\section{Discussions} \label{sec4}
\subsection{Height of magnetic reconnection: cause and consequence}

There are many eruptive phenomena in the solar atmosphere, from large solar flares to small-scale jets or brightenings. It is well known that magnetic reconnection plays a key role in the energy-releasing processes of these phenomena. In this sense, \citet{Shib99} proposed a unified model, i.e., plasmoid-induced magnetic reconnection, to explain all these phenomena. Along this line of thought, \citet{Chen99} proposed a unified model for solar flares, i.e., both two-ribbon flares and compact flares are produced by magnetic reconnection, and it is the height of reconnection that determines the morphological difference of the flares, i.e., magnetic reconnection at higher corona results in two-ribbon flares and magnetic reconnection at lower corona results in compact flares. Such a unified model was extended to other smaller eruptions. For example, it was shown by numerical simulations that magnetic reconnection in both the low corona and the chromosphere can produce microflares. However, the reconnection in the low corona can account for the microflares with soft X-ray jets, whereas the reconnection in the chromosphere is responsible for the microflares accompanied by H$\alpha$ or Ca brightenings \citep{Jian12}. 

\begin{figure}[htbp]
    \centering
    \includegraphics[width=0.7\textwidth]{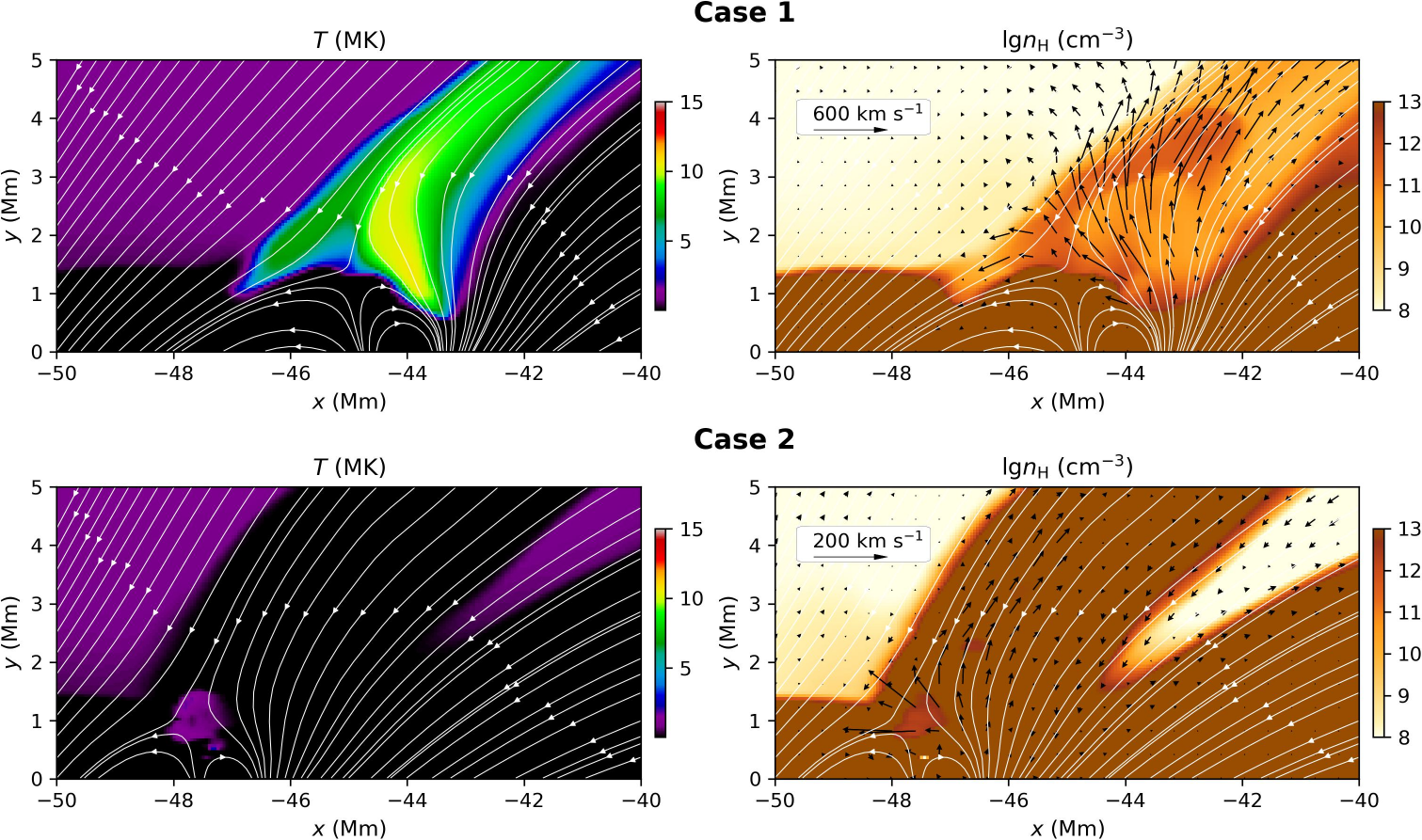}
    \caption{Distributions of temperature and density ({\it color scale}) around the magnetic reconnection site in the two cases, i.e., Case 1 at $t$=0.79 hr (top row) and Case 2 at $t$=0.67 hr (bottom row). Magnetic field is overlaid {\it white lines}.}
     \label{fig5}
\end{figure}

In this paper, we performed MHD numerical simulations of prominence formation due to magnetic reconneciton-induced localized heating in the chromosphere. By varying the location of the ephemeral magnetic flux, reconnection occurs either in the upper or lower chromosphere. When reconnection takes place in the upper chromosphere at an altitude of $\sim$1.47 Mm above the photosphere (Case 1), as shown in the top row of Figure \ref{fig5}, the reconnection outflow is heated to $\sim$10 MK and ejected to the corona as a hot jet with a velocity of $\sim$600 $\rm km\ s^{-1}$, close to the local Alfv\'en speed in the inflow region. This jet is primarily accelerated by the combined action of the Lorentz force and the gas pressure gradient. The jet travels along the magnetic field lines in the corona. As it overshoots the right shoulder of the dipped magnetic field structure, the jet is reflected at the right footpoint. The ensuing forward-propagating flows collide with the reflected plasma, facilitating the accumulation of hot material in the corona. Eventually, the magnetic dip becomes thermally unstable, and the hot plasma cools down and condenses to form a prominence. This scenario resembles the chromospheric evaporation–coronal condensation model of prominence formation.

In contrast, when magnetic reconnection occurs in the lower chromosphere at an altitude of $\sim$0.91 Mm (Case 2), as seen in the bottom row of Figure \ref{fig5}, only the local area around the reconnection site in the lower chromosphere is heated to $\sim$1 MK. The gas pressure gradient force, combined with magnetic tension, drives the cold upper chromosphere to move up as a cold surge. The cold surge with a temperature $\sim$70,000 K and densities around $10^{10.1}\ \rm cm^{-3}$ is accelerated to a velocity of $\sim$40 $\rm km\ s^{-1}$. Compared to the jet in Case 1, the surge in Case 2 exhibits higher densities, lower temperature, and smaller velocities. In this scenario, magnetic reconnection proceeds along a set of magnetic field lines with relatively shallower dips. Consequently, only part of the cold material remains trapped in the magnetic dip above the PIL, forming a suspended cold prominence. A fraction of the cold surge overshoots the right shoulder of the dipped magnetic field line, while some of the cold surge in the tail part drains back to the original footpoint as coronal rains. This process resembles the direct injection model for prominence formation.

It is seen that our 2D MHD numerical simulations confirmed the conclusion in our 1D hydrodynamic simulations, where the reconnection-related heating at different heights in the chromosphere was assumed phenomenologically. In this paper, the magnetic reconnection-related heating at different heights is realized more self-consistently compared to our 1D model \citep{Huan21}. When the preexisting magnetic field at the ephemeral flux region is a little more inclined, the reconnection site is located in the upper chromosphere; When the preexisting magnetic field is more vertical, the reconnection site is located in the lower chromosphere. In the real solar atmosphere, magnetic field is distributed as a canopy structure due to photospheric convective motions. As pointed out by \citet{Jian12}, when the emerging flux is slightly away from the photospheric magnetic elements, the ephemeral flux reconnects with the canopy field lines at the upper chromosphere. In this case, magnetic reconnection would evaporate the upper chromospheric plasma into the corona. Only after thermal instability can a prominence be formed. When the emerging flux is closer to the photospheric magnetic elements (often the boundary of supergranules), the ephemeral flux reconnects with the canopy field lines at a lower height. In this case, magnetic reconnection would inject cold chromospheric materials directly into the corona, forming a prominence described in the direct injection model. 

It is noted that the evaporation--condensation scenario and the direct injection scenario are not exclusive. For example, in Case 1, short injection of cold plasma precedes the evaporation process. Besides, compared to the simulation box, the current sheet of magnetic reconnection is too small for magnetic islands to be reproduced. It is expected that tearing mode instability would result in magnetic islands during the reconnection process as simulated by \citet{ni15}, \citet{zhao22}, and \citet{Lixh25}, which would lead to intermittent evaporation of hot plasma in Case 1 and intermittent injection of cold plasma in Case 2.

\subsection{Mass circulation in the low solar atmosphere}

Mass circulation is an important ingredient of the plasma dynamics between the corona and lower solar atmosphere, which is related to both coronal heating and condensation. Solar prominences, as well as coronal rains, are the important manifestations of the plasma circulation, where plasmas in the cold chromosphere are heated and evaporated to or pushed to the corona \citep{Anto12, Kohu19, liht23, kuce24}. While some of the hot coronal plasmas move out in the form of solar wind, other plasmas return to the solar surface in the form of coronal rains or prominence drainage. 

\begin{figure}[htbp]
	\centering
	\includegraphics[width=0.6\textwidth]{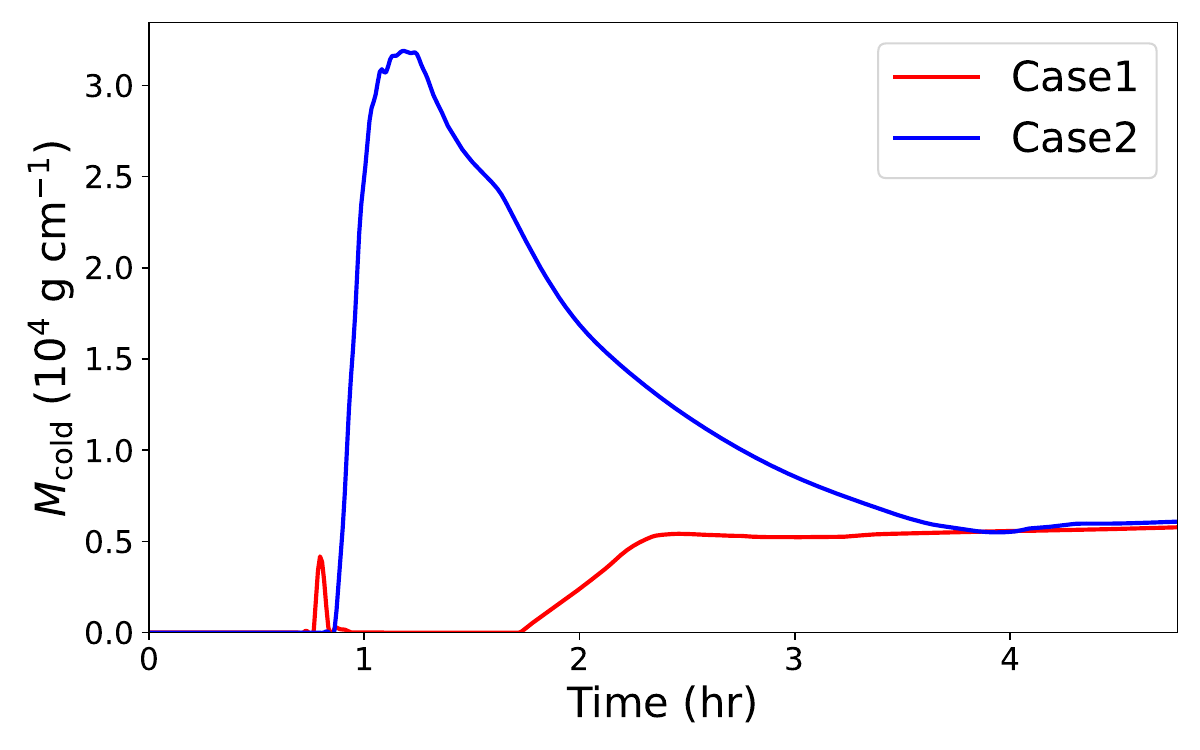}
	\caption{Evolution of the total prominence mass (per unit length in the $z$-direction) in the corona for Case 1 ({\it red line}) and Case 2 ({\it blue line}). }
	\label{fig6}
\end{figure}

To illustrate the plasma circulation during the prominence formation process in the two cases, we plot the evolution of the prominence mass (per unit length in the $z$-direction) in Figure \ref{fig6}, where all plasmas with temperature below 20,000 K inside the magnetic dip region are included. It is seen that in Case 1 ({\it red line}), there is a short episode of coronal condensation at around $t$=0.8 hr. Soon the cold condensation drains down to the solar surface completely due to strong dynamics associated with fast reconnection at the left footpoint of the field lines. At $t$=1.7 hr, condensation appears again and starts to grow linearly until $t$=2.3 hr. Thereafter, the total mass saturates at $0.41\times 10^4$ g cm$^{-1}$ as no more chromospheric evaporation fills into the corona from the ephemeral magnetic site. During the whole process, about $4.16\times 10^5$ g cm$^{-1}$ of hot mass is evaporated into the corona, i.e., almost about 99\% of the evaporated mass drains down to the solar surface. In Case 2 ({\it blue line}), cold chromospheric plasmas are injected into the magnetic dip at around $t$=0.84 hr. Compared to Case 1, more plasmas are accumulated in the magnetic dip region of the corona, reaching maximum at around $t$=1.2 hr. Gradually most of the cold plasmas drain down to the solar surface, and only 17\% of the maximum mass remains inside the magnetic dips.

It is noticeable that in both cases the majority of the plasma supplied to the corona eventually drains down to the solar surface, with only a small fraction remaining confined within the coronal region. Unlike the 3D simulations \citep{Xiac16}, where the total prominence mass can be calculated, our 2D simulations can only provide linear density. Nevertheless, our results are qualitatively consistent with the observations of \citet{Liuw12}, who reported that approximately 96\% of the total condensation drains back to the solar surface within roughly one day. Such mass circulation is expected to play a key role in redistributing and mixing elemental abundances between the corona and chromosphere \citep{ng24}.

To summarize, in this paper we performed 2D MHD simulations of the magnetic reconnection between emerging ephemeral flux and the preexisting magnetic field, with the consideration of gravity, thermal conduction, radiative loss, heating and anomalous resistivity. It is revealed that magnetic reconnection happens either in the upper chromosphere or in the lower chromosphere, depending on the emerging site. In the former case, upper chromospheric plasma is heated to coronal temperatures and is evaporated into the corona. The resulting dense hot plasma in the corona cools down due to thermal instability to form a prominence in a way described by the traditional evaporation--condensation model. In the latter case, only the lower chromospheric plasma is heated, and the cold chromospheric plasma in the upper chromosphere is pushed by the gas pressure and Lorentz force so as to be injected to the corona. Part of the injected cold plasma is trapped in magnetic dips, forming a prominence in a way described by the traditional direct injection model. Our simulation results verified our unified model for prominence formation, i.e., the traditional evaporation--condensation model and the direct injection model can be unified in one framework, and the different manifestations of the two forming mechanisms are simply due to the difference of reconnection height in the chromosphere. 

It is pointed out that in this paper, the plasma is considered as a single fluid. However, the chromosphere and photosphere are partially ionized. In this sense, not only should the ionization potential be taken in account in the energy equation \citep{Chen01}, but also the low solar atmosphere should be treated with the multi-fluid model, as done by \citet{pope23}. Both factors might influence the heating process in the chromosphere, hence modulate the evaporation rate or injection rate in our two cases. Besides, extending our unified model from 2D to 3D would also be valuable as more fine structures of the prominences formed in the two scenarios can be revealed. In particular, in the 3D case, the initial magnetic configurations can be either a magnetic arcade or flux rope \citep{ouya17}. It would be interesting to investigate how the initial magnetic configuration influences the appearance of prominences formed in the two scenarios of our unified model.
]

\bibliography{ms}{}
\bibliographystyle{aasjournal}

\end{document}